%% file: main.tex
\documentclass[DIV=12,numbers=noenddot]{scrartcl}

\usepackage[utf8]{inputenc}
\usepackage[T1]{fontenc}
\usepackage{lmodern}
\usepackage{amsmath,amssymb}
\usepackage[usenames]{xcolor}
\usepackage{xspace}
\usepackage{booktabs}
\usepackage{authblk}
\usepackage[font=small,labelfont=bf,format=plain,margin=0.05\textwidth]{caption}
\usepackage{subcaption}
\usepackage{tabularx}
\usepackage{listings}

\usepackage[frozencache]{minted}

\newcommand{\mytitle}{Phenomenology of the Higgs sector from Reduction of Couplings in the Type-II 2HDM}
\usepackage{relsize}
\newcommand{\smallunderscore}{\textscale{.5}{\textunderscore}}

\usepackage[
  pdftitle={\mytitle},
  pdfauthor={Wojciech Kotlarski, Gregory Patellis},
  pdfkeywords={},
  bookmarks=true, linktocpage, colorlinks=true, allbordercolors=white,
  allcolors=blue]{hyperref}
\usepackage{orcidlink} % must come AFTER hyperref

\usepackage[
  backend=biber,
  bibencoding=utf8,
  sorting=none,
  url=true,
  doi=false,
  style=phys,
  biblabel=brackets,
  subentry=false,
  eprint=true,
  maxbibnames=5
]{biblatex}
\title{\mytitle}
\date{}
\author[a]{Wojciech Kotlarski\orcidlink{0000-0002-1191-6343}\thanks{\href{mailto:wojciech.kotlarski@ncbj.gov.pl}{wojciech.kotlarski@ncbj.gov.pl}}}
\affil[a]{National Centre for Nuclear Research, Pasteura 7, 02-093 Warsaw, Poland}
\author[b]{Gregory Patellis\orcidlink{0000-0002-5053-3994}\thanks{\href{mailto:grigorios.patellis@tecnico.ulisboa.pt}{grigorios.patellis@tecnico.ulisboa.pt}}}
\affil[b]{Centro de Física Teórica de Partículas - CFTP, Departamento de Física, Instituto Superior Técnico, Universidade de Lisboa,
Avenida Rovisco Pais 1, 1049-001 Lisboa, Portugal}

\newcommand{\FS}{\texttt{Flex\-ib\-le\-SUSY\@}\xspace}
\newcommand{\sarah}{\texttt{SARAH}\@\xspace}
\newcommand{\HB}{\texttt{Higgs\-Bounds}\@\xspace}
\newcommand{\HS}{\texttt{Higgs\-Signals}\@\xspace}
\newcommand{\HT}{\texttt{Higgs\-Tools}\@\xspace}
\DeclareMathOperator{\hc}{h.c.}

\begin{document}
\maketitle
\input{tex/abstract}

%\tableofcontents

\input{tex/introduction}
\input{tex/model}

\input{tex/results}
\input{tex/conclusions}
\input{tex/acknowledgements}

\appendix
\input{tex/appendix}

\printbibliography[title={References}]
\addcontentsline{toc}{section}{References}

\end{document}

%% file: tex/abstract.tex
The idea of \textit{reduction of couplings} provides a systematic procedure to search for relations among seemingly unrelated parameters of a renormalizable theory. 
As a consequence, such reduced theories exhibit more constrained parameter spaces.
Motivated by this, in this work we perform a precise phenomenological analysis of the Higgs sector of a version of the Type-II 2HDM on which the idea of reduction of couplings has been applied.
We compute Higgs boson masses and decay widths and confront them with current experimental measurements.
Compared to the previous study, apart from the inclusion of actual experimental constraints on production and decay rates of Higgs bosons, we also included all model parameters in the RGE running as well as one- and two-loop threshold corrections and two-loop RGE running from the high scale.
Furthermore, Higgs boson masses are now computed at the one-loop level.
Top quark mass, which is an important prediction of the reduction framework, is now evaluated including previously missing sub-leading one-loop contributions, with an addition of up to four-loop pure-QCD corrections.

%% file: tex/introduction.tex
\section{Introduction}

Despite the many successes of the Standard Model (SM) in describing elementary particles and their interactions, one of its problems is the arbitrariness of its parameter space. This issue is deeply related to the infinities that emerge at the quantum level. Renormalization indeed removes those infinities, but at the cost of introduction of counter terms. This leaves the `cured' parameters free and they are expected to be fixed by experimental measurements.

One of the most popular ways to constrain this parameter space is the introduction of extra symmetries, whether in the form of an additional (local or global) symmetry or as a covering gauge group. However, such models often present new complications stemming from the number of new independent parameters, as extra degrees of freedom are usually necessary in order for the models to become realistic (e.g. to break the gauge symmetry in the case of a Grand Unified Theory - GUT).

As an alternative, the idea of  \textit{reduction of couplings} (RoC) was proposed in \cite{Zimmermann:1984sx} and \cite{Oehme:1984yy,Oehme:1985jy}. The RoC technique reduces the freedom of a renormalizable theory's parameter space by systematically expressing either all or a number of its seemingly independent dimensionless parameters in terms of a single, independent, `primary' coupling. This gives the theory significant predictive power. In its original version, such relations among the reduced parameters and the primary coupling are additionally renormalization group invariant (RGI).

RoC was initially applied to the SM \cite{Kubo:1985up,Kubo:1988zu} providing promising (at the time) predictions for the top quark and Higgs boson masses.
The discovery of the top quark (and subsequently of the Higgs boson) ruled out the attempt, as the predicted masses turned out to be too light. However, this spurred interest in a number of reduced theories that are based on SM extensions (i.e. \cite{Kubo:1994bj,Heinemeyer:2007tz}). The current status of some of these models can be found in \cite{Heinemeyer:2020ftk,Heinemeyer:2020nzi}.

In the modern version of RoC, the method is applied either above or at a specific energy scale. Below that boundary scale the renormalization group equations (RGEs) of the theory run as usual, using the reduction relations among the reduced parameters and the primary coupling as boundary conditions. In the case these relations are RGI above the boundary scale a specific covering theory is explicitly considered (typically a GUT), while when the reduction is done at the boundary the gauge structure or additional field content of the covering theory can be left implicit, as the only requirement is that they validate the reduction relations at the boundary. An interesting feature of the latter case is that RoC can naturally result in symmetries of the Lagrangian without the need to impose them on the theory.

In the previous work \cite{Pech:2023bjm} the RoC method was applied at one-loop level to a version of the Type-II Two Higgs Doublet Model (2HDM), where all of the parameters of the Higgs potential and the Yukawa couplings are real. The relations among the quartic Higgs couplings, the top Yukawa coupling and the \textit{primary} strong coupling $g_s$ (with corrections from the other two SM gauge couplings, $g'$ and $g$), hold at the boundary scale $\mu\sim 10^7$ GeV and serve as boundary conditions for the RGE running down to the electroweak scale. By fixing the value of $\tan\beta$ to around 2 the obtained one-loop pole top quark mass is close to the current experimental limits, while five sets of solutions give a viable tree-level (light) Higgs boson mass. Notably, this was the first non-supersymmetric reduced model that featured a viable mass spectrum.

In this work we extend on the previous analysis and perform a state-of-the-art phenomenological analysis of the Higgs sector of this model.
We employ publicly available tools allowing for very precise evaluation of Higgs physics observables and their confrontation with experimental data.
With this, we prove the viability of the RoC scenario in the 2HDM context.

The paper is structured as follows.
In Sec.~\ref{sec:model} we review the real Type-II 2HDM with boundary conditions coming from the reduction of couplings and briefly discuss its implementation in \FS spectrum-generator generator.
In Sec.~\ref{sec:results} we describe the setup of our analysis and perform a detailed phenomenological study of the set of parameter points obtained from RoC.
Finally, we summarize our findings in Sec.~\ref{sec:conclusions}.
We relegate finer details related to the implementation of this model in \FS to Appendix~\ref{app:fs}.

%% file: tex/model.tex
\section{The Reduced Real Type-II 2HDM} \label{sec:model}

The idea of RoC is to search for relations between a dimensionless parameter that stays independent and is called the `primary' coupling and the rest (or some) of the dimensionless parameters of a theory. Assuming such relations exist among an $A$ number of couplings, in order to find them one has to solve the so called reduction relations (REs) \cite{Zimmermann:1984sx,Oehme:1984yy,Oehme:1985jy},
\begin{equation}
\beta_{g} \,\frac{d g_{a}}{d g} =\beta_{a}~,~~~~~~~~a=1,\ldots,A-1,
\label{redeq}
\end{equation}
where $g$ and $g_a$ are the  primary
coupling and the reduced couplings, respectively, and $\beta_{g}$, $\beta_{g_a}$ their corresponding $\beta$-functions.
In order not to have the integrations constants that usually come with solving ordinary differential equations such as the REs, the solutions should come in the form of power series,
\begin{equation}
g_{a} = \sum_{n}\rho_{a}^{(n)}\,g^{2n(+1)}~,
\label{powerser}
\end{equation}
which also preserve perturbative renormalizability and are called reduction solutions (RSs). While in the original version the RSs are RGI and RoC holds in the entire energy regime of the theory, in later applications of the method \cite{Mondragon:2013aea,Heinemeyer:2017gsv,Pech:2023bjm} the RSs are found at a UV boundary scale of the theory and are used as boundary conditions for the RGE running of the reduced parameters. Above the boundary a covering theory is assumed, for which the RSs should also hold at the boundary or can even be RGI.

On a practical level, reduction provides boundary conditions for a BSM model at the high scale, similar to how a Constrained Minimal Supersymmetric Standard Model provides boundary conditions for a minimally supersymmetrized Standard Model.

For the case of the Type-II 2HDM, as in most reduced theories, the `primary' coupling is chosen to be the strong gauge coupling, $g_s$. The $SU_L(2)$ and $U_Y(1)$ gauge couplings, $g$ and $g'$, are treated as corrections to the reduction. This sets the boundary scale to $\mu = 10^7$ GeV since above that the $g$ cannot be treated as a correction only any more (see \cite{Pech:2023bjm} for details).
Since a complete reduction over all the dimensionless couplings of the theory is unrealistic, we only consider a reduction on the third fermionic generation.  The bottom quark and tau lepton Yukawa couplings are much smaller than the top quark one and thus are omitted for simplicity, but could be taken into account as threshold corrections.

We start from the reduction of the top Yukawa coupling $y_t$, since its 1-loop $\beta$-function does not contain any of the quartic scalar couplings and thus $y_t$ can be reduced independently. The RS that can lead to a viable top quark mass is found to be
\begin{equation}
    y_t~=~0.471g_s-0.119g+1.228g'~.\label{eq:RSyt}
\end{equation}
This relation is used as a boundary condition at $\mu=10^7$GeV for the RG running of $y_t$ down to the EW scale.
The final value of top-quark mass depends on the value of the ratio of vacuum expectation values of the two Higgs doublets, which is a free parameter of the reduced theory (since it is not related to dimensionless parameters of the theory).
Ref.~\cite{Pech:2023bjm} identified that $\tan \beta$ should be around 2.
We will refine this prediction in this analysis.

Reduction of the Higgs potential parameters $\lambda$ proceeds analogously.
We start with the most general renormalizable scalar potential with two Higgs doublets $\Phi_1$ and $\Phi_2$, written as \cite{Wu:1994ja,Davidson:2005cw,Branco:2011iw}
\begin{align}\label{eq:potential}
    V_h=&~~~m_{11}^2\Phi_1^{\dagger}\Phi_1+m_{2}^2\Phi_2^{\dagger}\Phi_2-\Big(m_{12}^2\Phi_1^{\dagger}\Phi_2+ \hc \Big)\nonumber\\
    &+\frac{1}{2}\lambda_1\Big(\Phi_1^{\dagger}\Phi_1\Big)^2+\frac{1}{2}\lambda_2\Big(\Phi_2^{\dagger}\Phi_2\Big)^2+\lambda_3\Big(\Phi_1^{\dagger}\Phi_1\Big)\Big(\Phi_2^{\dagger}\Phi_2\Big)+\lambda_4\Big(\Phi_1^{\dagger}\Phi_2\Big)\Big(\Phi_2^{\dagger}\Phi_1\Big)\nonumber\\
    &+\Bigg[\frac{1}{2}\lambda_5\Big(\Phi_1^{\dagger}\Phi_2\Big)^2+\lambda_6\Big(\Phi_1^{\dagger}\Phi_1\Big)\Big(\Phi_1^{\dagger}\Phi_2\Big)+\lambda_7\Big(\Phi_2^{\dagger}\Phi_2\Big)\Big(\Phi_1^{\dagger}\Phi_2\Big)+ \hc\Bigg],
\end{align}
where all parameters are taken to be real. We work in the Type-II scenario, thus up-quarks couple with $\Phi_2$, while down-quarks and charged leptons couple to $\Phi_1$. We also assume that vacuum expectation values $v_1$ and $v_2$ of doublets $\Phi_1$ and $\Phi_2$ are real. Since the above-mentioned RoC works among dimensionless parameters, the squared scalar mass parameters $m_{11}^2, m_{22}^2$ and $m_{12}^2$ will remain unconstrained as is the case with $\tan \beta \equiv v_2/v_1$.

Using 1-loop RGEs we obtain sets of RSs that serve as boundary conditions for $\lambda$'s
\begin{align}
    \lambda_i = p_i g_s^2 + q_i g^2 + r_i g'^2~,\label{eq:lambda_bc}
\end{align}
where $p$, $q$ and $r$ are concrete numbers.
In total, we obtain a set of 509 solutions.
Most of them are not phenomenologically viable, but at least some were able to predict correct (within the precision of analysis in Ref.~\cite{Pech:2023bjm}) SM-like Higgs and top masses.
The refinement of these predictions is the main goal of this work.

%% file: tex/results.tex
\section{Phenomenological analysis} \label{sec:results}

For numerical analysis, we implement the Real Type-II 2HDM with the boundary conditions in Eqs.~\eqref{eq:RSyt} and \eqref{eq:lambda_bc} in the \FS spectrum-generator generator \cite{Athron:2014yba,Athron:2017fvs}.
Embedding the model within \FS allows to refine three aspect of the previous analysis.

First, the SM-like parameters are now determined via matching to the Standard Model at $\mu=m_Z$ using 2-loop threshold corrections, as is customary in state-of-the-art analyses of BSM models.
Moreover, all model parameters are now included in the RGE running from $10^7$ GeV to EW scale or vice versa.
This introduces, for example, dependence of $\lambda$'s on sub-leading contributions from non-top Yukawas.
In addition, the running from the high scale is now performed using 2-loop RGEs \cite{Staub:2010jh}.

Second, \FS provides the possibility to calculate top quark mass, which is a prediction of RoC 2DMII, with full NLO BSM corrections and up to 4-loop pure QCD ones \cite{Athron:2017fvs,Avdeev:1997sz,Bednyakov:2002sf,Melnikov:2000qh,Chetyrkin:1999qi,Martin:2016xsp}.
In order to satisfy $m_t = 172.57 \pm 0.29$ GeV \cite{PhysRevD.110.030001}, the ratio of the two Higgs vevs is fixed to $\tan\beta=1.85$, with a very small allowed window as opposed to the value of $2.2 \pm 0.1$ predicted in \cite{Pech:2023bjm}.
We quantify these findings further at the end of this section.

Finally, we assess the compatibility of the Higgs sector with multitude of experimental data, not just the value of the SM-like Higgs boson mass.
We compute Higgs boson mass at one-loop level and compute Higgs decays using \texttt{FlexibleDecay} module of \FS \cite{Athron:2021kve}.
The properties of predicted Higgs bosons are checked against experimental data using \FS interface to \HT \cite{Bahl:2022igd}.\footnote{\FS interface to \HT and \texttt{Lilith} is available since \FS v2.9.0, which can be downloaded from \href{https://github.com/FlexibleSUSY/FlexibleSUSY/archive/refs/tags/v2.9.0.tar.gz}{https://github.com/FlexibleSUSY/FlexibleSUSY/archive/refs/tags/v2.9.0.tar.gz}.}
The scan over the set of RoC solutions is performed using the \texttt{PySLHA} \cite{Buckley:2013jua} package.

In the implementation, dimensionful parameters $m_{11}^2$ and $m_{22}^2$ in Eq.~\eqref{eq:potential} are eliminated through tadpole equations.
For consistency with Ref.~\cite{Pech:2023bjm}, we use tree-level mass of the physical pseudoscalar Higgs $m_A$ as an input parameter.
We use it to replace remaining dimensionful potential parameter $m_{12}$ in Eq.~\eqref{eq:potential} at the scale $\mu=125$~GeV, at which the mass spectrum is evaluated, as
\begin{align}
\label{eq:mA}
m_{12}^2 = \lambda_5 v_1 v_2 + \frac{1}{2}\left(\lambda_6 v_1^2 + \lambda_7 v_2^2\right) + \frac{v_1 v_2}{v_1^2 + v_2^2} m_A^2~.
\end{align}
The boundary conditions for $y_t$ and $\lambda$'s are imposed as explained in Sec.~\ref{sec:model}, at the scale $\mu = 10^7$ GeV.
We provide a more detailed explanation of the implementation in Appendix~\ref{app:fs}.

\begin{figure}
    \centering
    \includegraphics[width=0.6\linewidth]{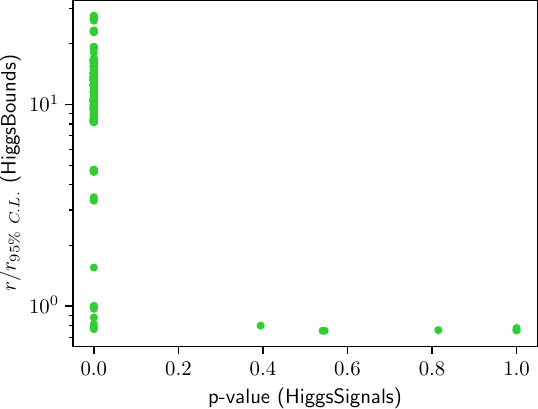}
    \caption{The \HS p-value and the maximal (over the checked Higgs states) observed ratio to 95\% C.L. exclusion from \HB for a set of 509 points coming from RoC. Few points are clustered at exactly p-value = 1.}
    \label{fig:hs_vs_hb}
\end{figure}

In Fig.~\ref{fig:hs_vs_hb} we show p-values reported by \HS and max
ratios of predicted cross section and the experimental limit at the 95\% confidence level reported by \HB for all of the 509 points obtained from RoC.
In the analysis, we use \HS database v1.2 and \HB database v1.6.
Most points are excluded by \HS.
There are however 9 point allowed by \HB and with \HS p-value $\geq 0.05$, with 5 points with p-value exactly equal to 1.
All these points also pass basic theoretical requirements like boundedness-from-below and vacuum stability \cite{Branco:2011iw}, and unitarity \cite{Goodsell:2018tti}.
This confirms the viability of RoC as a guiding principle for BSM model building.

In \FS interface to \HT we include a 3\% uncertainty on the Higgs mass, meaning that in principle point with mass between around 122 and 128 could have a \HS p-value of 1.
As an benchmark we provide here the best point, with Higgs mass closest to 125.2 GeV, which has the following parameters (for ease of read we have round down the actual parameters to 3 digits of precision below)\footnote{The \texttt{SLHA}~\cite{Skands:2003cj,Allanach:2008qq} input file for that point is attached to the \texttt{arXiv} version of this work.}
\begin{align}
    \lambda_1 =& -0.748 g_s^2 + 2.74 g^2 + 5.86 g'^2 \label{eq:best_l1}\\
    \lambda_2 =& -1.16 g_s^2 + 3.77 g^2 -4.11 g'^2 \\
    \lambda_3 =& -0.970 g_s^2 + 3.35 g^2 + 6.09 g'^2 \\
    \lambda_4 =& -7.86 g'^2 \\
    \lambda_5 =& \lambda_6 = \lambda_7 = 0 \label{eq:best_l7}
\end{align}
giving $m_t = 172.3$ GeV and $m_h = 125.1$ GeV, $m_H = 996$ GeV, $m_A = 991$ GeV, $m_{H^\pm} = 1002$ GeV.
\HB ratio for this point is 0.76, with the most constraining analysis being \cite{CMS-PAS-HIG-12-045}.
We note that \FS interface to \HT only checks Higgs bosons with masses below around 650 GeV, meaning that non-SM states are not checked for this parameter point.
However, since they are relatively heavy, they are not expected to be constrained by currently available data; neither by Higgs constraints nor by the electroweak precision tests nor by flavour physics.
For example, due to the mass degeneracy, one expects that Peskin-Takeuchi \cite{Peskin:1990zt,Peskin:1991sw} parameters for that point are small, $S \approx T \approx 0$.
The predicted W-boson mass, based on the full 1-loop  calculation including higher-order SM corrections \cite{Athron:2022isz}, of $80.355$~GeV is in agreement with the PDG value of $80.3692 \pm 0.0133$~GeV \cite{PhysRevD.110.030001}.
Finally, the charged Higgs boson mass is also well above the limit from b-physics, which is currently set around 580~GeV in the Type-II 2HDM for $\tan \beta \gtrsim 2$ \cite{Misiak:2017bgg}.
\begin{figure}
    \centering
    \includegraphics[width=0.6\linewidth]{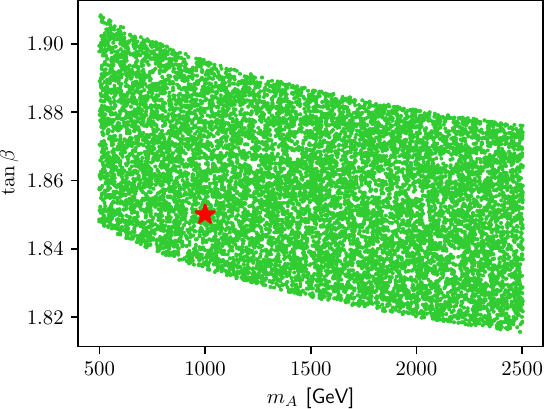}
    \caption{Variation of $m_A$ and $\tan \beta$ around the best-fit point from Eqs.~\eqref{eq:best_l1}--\eqref{eq:best_l7}. Shown points are allowed at 95\% C.L. by top mass, \HS and \HB.}
    \label{fig:mA_vs_tanb}
\end{figure}

We now turn to the dependence of the benchmark point on the free parameters of the reduction procedure: the $\tan \beta$ and the mass of the pseudoscalar $m_A$.
Fig.~\ref{fig:mA_vs_tanb} shows point allowed at 95\% C.L. by top mass, \HS and \HB around the best-fit point.
Of those three constraints the allowed region, as expected, turns out to be constrained solely by top mass and \HB constraints as $m_A$ and $\tan \beta$ in this range do not influence SM-like Higgs to any extent.
Requiring a correct top-quark mass prediction results in a very narrow allowed window for $\tan \beta$, with allowed variation of around $\pm 0.04$.
This is a substantial refinement over the constraint obtained in Ref.~\cite{Pech:2023bjm}, where $\tan \beta$ was narrowed down only to $\pm 0.1$.

%% file: tex/conclusions.tex
\section{Conclusions} \label{sec:conclusions}

\textit{Reduction of couplings} is a powerful principle that allows to constrain parameter spaces of Beyond the Standard Model theories by relating their seemingly independent parameters.
Initially, the RoC principle has been applied in the context of the pure SM and supersymmetric models but recently has seen its first uses in non-supersymmetric cases.

One of the most popular non-supersymmetric BSM extensions of the Higgs sector is the 2 Higgs Doublet Model.
The 2HDM extends the SM by introducing a second Higgs doublet, extending the scalar sector of the theory and offering a framework to address phenomena beyond the SM.
In the context of RoC, Ref.~\cite{Pech:2023bjm} showed that the Reduced Type-II 2HDM is able to provide a Higgs boson with a tree-level mass of roughly the appropriate value as well as the top-quark mass close to the experimental measurement as predicted based on a partial one-loop calculation.

Motivated by these promising results, in this work we perform a more realistic analysis of this scenario.
We created a dedicated spectrum generator for the Type-II 2HDM with RoC boundary conditions within \FS spectrum-generator generator.
This allowed us to include RGE running of previously fixed parameters like for example $m_{12}^2$ and $\tan \beta$, a two-loop running between the high and the electroweak scale and a proper matching of SM-like parameters at the electroweak threshold.
The Higgs boson masses are now computed at the one-loop level, with top-quark masses computed including previously missing parts of one-loop BSM contributions and up to four-loop QCD ones.
Finally, we confront our findings with experimental measurements by using the \FS interface to \HT to confirm the validity of this model's Higgs sector.

Out of the sets of 509 reduced solutions we identify points with Higgs mass of around 125 GeV and a correct top-quark mass, with Higgs sector properties in agreement with current experimental data (see Fig~\ref{fig:hs_vs_hb}).
The best of such points is provided as a benchmark.

While reduction does not tell us anything about dimensionful parameters like the mass of pseudoscalar Higgs  or $\tan \beta$ (which we refer to as dimensionful since it describes individual vevs of two Higgs doublets where the total magnitude of $v_1^2 + v_2^2$ is fixed by the electroweak data) we show that for a tree-level mass of $m_A (\mu=125~\text{GeV})$ between 500 and 2500 GeV, $\tan \beta(\mu=m_Z)$ has to be between 1.82 and 1.91 (the precise window of $\tan \beta$ depends on a specific value of $m_A$, see Fig.~\ref{fig:mA_vs_tanb}) to allow for a 95\% C.L. agreement with Higgs data and the top-quark mass measurement.
This is a significant refinement over the previously found value of $\tan \beta = 2.2 \pm 0.1$.

Direct probe of RoC predictions in the 2HDM requires measurement of Higgs self interactions.
While this is currently not possible, future experimets, like the FCC, will provide for that an ample oportunity.
Measurement of triple-Higgs couplings, either from double Higgs production or from Higgs to Higgs decay after the discovery of heavy Higgses, will allow to verify the specific set of potential parameters $\lambda_i$ predicted by the reduction.

%% file: tex/acknowledgements.tex
\section*{Acknowledgements}

The authors thank Alexander Voigt for his help with \FS and Myriam Mondragón and George Zoupanos for helpful discussions on the theoretical aspects of reduction.

WK is supported by the National Science Centre (Poland) grant 2022/\allowbreak47/\allowbreak D/\allowbreak ST2/\allowbreak03087.
GP is supported by the Portuguese Funda\c{c}\~{a}o para a Ci\^{e}ncia e Tecnologia (FCT) under Contracts UIDB/00777/2020 and UIDP/00777/2020. 
These projects are partially funded through POCTI (FEDER), COMPETE, QREN and the EU.
GP has a postdoctoral fellowship in the framework of UIDP/00777/2020 with reference BL154/2022\smallunderscore IST\smallunderscore ID.

%% file: tex/appendix.tex
\section{Implementation of RoC boundary conditions in \FS}
\label{app:fs}

Type-II 2HDM with RoC boundary conditions is implemented in \FS as a two-scale BSM model.
The high-scale, controlled by the \texttt{Qin} variable, is fixed to a value of $10^7$ GeV in the \texttt{SLHA} input file while the scale at which we compute the mass spectrum is set to \texttt{Qout} = 125 GeV.
The RoC boundary conditions are applied at the \texttt{Qin} scale
\begin{minted}[breaklines,linenos,frame=none,fontsize=\small]{mathematica}
HighScaleFirstGuess = Qin;    
HighScale = Qin;    
HighScaleInput = {    
    {Yu[3,3], -(ptIN*g3 + qtIN*g2 + rtIN*GUTNormalization[g1]*g1)},    
    {Lambda1, 1/2*(p1IN*g3^2 + q1IN*g2^2 + r1IN*(GUTNormalization[g1]*g1)^2)},    
    {Lambda2, 1/2*(p2IN*g3^2 + q2IN*g2^2 + r2IN*(GUTNormalization[g1]*g1)^2)},    
    {Lambda3, p3IN*g3^2 + q3IN*g2^2 + r3IN*(GUTNormalization[g1]*g1)^2},    
    {Lambda4, p4IN*g3^2 + q4IN*g2^2 + r4IN*(GUTNormalization[g1]*g1)^2},    
    {Lambda5, p5IN*g3^2 + q5IN*g2^2 + r5IN*(GUTNormalization[g1]*g1)^2},    
    {Lambda6, p6IN*g3^2 + q6IN*g2^2 + r6IN*(GUTNormalization[g1]*g1)^2},    
    {Lambda7, p7IN*g3^2 + q7IN*g2^2 + r7IN*(GUTNormalization[g1]*g1)^2}    
}; 
\end{minted}
The built-in \FS's Type-II 2HDM \sarah model file uses different convention for the Higgs potential; hence we internally rescale boundary values for $\lambda_{1,2}$ given by Eq.~\eqref{eq:lambda_bc} by a factor of $\tfrac{1}{2}$.
Additionally, $\lambda_5$ should be conjugated but since in the current analysis this parameter is real it is of no consequence to this discussion.

$m_{12}^2$ is eliminated in favour of a tree-level $m_A$ mass (see. Eq.~\ref{eq:mA}) at the \texttt{Qout} scale (\FS refers to his scale as \texttt{SUSYScale} even in non-SUSY models)
\begin{minted}[breaklines,linenos,frame=none,fontsize=\small]{mathematica}
SUSYScaleFirstGuess = Qout;    
SUSYScale = Qout;    
SUSYScaleInput = {    
   {M122, (Lambda6*v1^2)/2 + Lambda5*v1*v2 + (Lambda7*v2^2)/2 + (mAIN^2*v1*v2)/(v1^2+v2^2)}    
}; 
\end{minted}
Non-top quark Yukawas as well as vacuum expectations values are matched to the SM as usual at the $Z$-boson mass scale
\begin{minted}[breaklines,linenos,frame=none,fontsize=\small]{mathematica}
LowScaleInput = {    
   {v1, 2 MZMSbar/Sqrt[GUTNormalization[g1]^2 g1^2 + g2^2] Cos[ArcTan[TanBeta]]},    
   {v2, 2 MZMSbar/Sqrt[GUTNormalization[g1]^2 g1^2 + g2^2] Sin[ArcTan[TanBeta]]},
   {Yu[1,1],  -Sqrt[2] upQuarksDRbar[1,1]/v2},    
   {Yu[2,2],  -Sqrt[2] upQuarksDRbar[2,2]/v2},    
   {Yd, Automatic},    
   {Ye, Automatic}
};
\end{minted}
while the top-quark Yukawa \texttt{Yu[3,3]} is already fixed at the \texttt{HighScale}.

The complete \FS model file is attached to the \texttt{arXiv} version of this work.